\documentclass[acmsmall]{acmart}

\usepackage{algorithmic}
\usepackage{graphicx}
\usepackage{textcomp}
\usepackage[table]{xcolor}
\usepackage{balance} 
\usepackage{url} 
\usepackage{booktabs}
\usepackage{enumitem}
\usepackage{multirow}
\usepackage{subcaption} %
\usepackage{listings}
\usepackage{color}
\usepackage{threeparttable}
\usepackage{array}
\usepackage{diagbox}
\usepackage{graphicx}
\usepackage{verbatim}
\usepackage{url}
\usepackage{graphicx}
\usepackage{multirow}
\usepackage{multicol}
\usepackage{hhline}
\usepackage{comment}
\usepackage{float}
\usepackage{array,booktabs}
\usepackage{marvosym} %
\usepackage{pifont}
\usepackage{tcolorbox}
\usepackage{bm}
\usepackage[ruled,linesnumbered]{algorithm2e}
\usepackage{wrapfig}

\newcommand{\tabincell}[2]{\begin{tabular}{@{}#1@{}}#2\end{tabular}}

\newcommand{\find}[1]{
\begin{tcolorbox}[leftrule=0.5mm,rightrule=0.5mm, toprule=0.5mm,bottomrule=0.5mm,left=2pt,right=2pt,top=2pt,bottom=2pt]%
\em #1
\end{tcolorbox}
}

\newcommand{\todo}[1]{\textcolor{black}{#1}}
\newcommand{\todoa}[1]{\textcolor{black}{#1}}
\newcommand{\camera}[1]{\textcolor{black}{#1}}

\newcommand{\sir}{SVR\xspace}
\newcommand{\sirs}{SVRs\xspace}
\newcommand{\datasetsize}{6,446\xspace}
\newcommand{\datasetsizeproject}{1,986\xspace}

\newcommand{\proEVA}{{\sc proEVA}\xspace}

\newcommand{\appname}{{\sc EAVA}\xspace}
\newcommand{\appnamebold}{{\sc \textbf{EAVA}}\xspace}

\setcopyright{cc}
\setcctype{by}
\acmDOI{10.1145/3832188}
\acmYear{2026}
\acmJournal{PACMSE}
\acmVolume{3}
\acmNumber{ISSTA}
\acmArticle{ISSTA097}
\acmMonth{10}
\acmSubmissionID{issta26main-p882-p}
\received{2026-01-30}
\received[accepted]{2026-04-16}

\begin{document}

\title{Answer Is Cheap, Show Me the Evidence! Augmenting Automated Vulnerability Assessment with Evidence}

\author{Shengyi Pan}
\orcid{0000-0001-9471-9638}
\affiliation{%
  \institution{Zhejiang University}
  \city{Hangzhou}
  \country{China}
}
\email{shengyi.pan@zju.edu.cn}

\author{Zelong Zheng}
\orcid{0009-0006-2372-6586}
\affiliation{%
  \institution{Zhejiang University}
  \city{Hangzhou}
  \country{China}
}
\email{zelongzheng@zju.edu.cn}

\author{Jiayuan Zhou}
\orcid{0000-0002-5181-3146}
\affiliation{%
  \institution{Huawei}
  \city{Waterloo}
  \country{Canada}
}
\email{jiayuan.zhou1@huawei.com}

\author{Xing Hu}
\orcid{0000-0003-0093-3292}
\affiliation{%
  \institution{Zhejiang University}
  \city{Hangzhou}
  \country{China}
}
\email{xinghu@zju.edu.cn}

\author{Xin Xia}
\orcid{0000-0002-6302-3256}
\affiliation{%
  \institution{Zhejiang University}
  \city{Hangzhou}
  \country{China}
}
\affiliation{%
  \institution{Hangzhou High-Tech Zone (Binjiang) Institute of Blockchain and Data Security}
  \city{Hangzhou}
  \country{China}
}
\email{xin.xia@acm.org}
\authornote{Corresponding Author}

\author{Shanping Li}
\orcid{0000-0003-2615-9792}
\affiliation{%
  \institution{Zhejiang University}
  \city{Hangzhou}
  \country{China}
}
\email{shan@zju.edu.cn}

\begin{abstract}
Software Vulnerability (SV) assessment is a vital phase in SV management, which characterizes discovered SVs to locate hot spots and prioritize their remediation.
To reduce the overhead and latency of manual assessment, 
prior works have explored automatically predicting assessment results from SV reports (\sirs).
However, existing approaches fail to process the information conveyed by the rich text content (e.g., screenshots and code snippets) embedded in \sirs and miss information about vulnerable projects.
More importantly, they primarily focus on assessment accuracy while neglecting to provide explanations or evidence supporting their predictions. 
As a result, these approaches remain impractical in real-world settings, where imperfect accuracy necessitates manual validation.
LLMs offer a promising opportunity to address this limitation by performing SV assessment while simultaneously providing supporting evidence.
Nevertheless, our extensive evaluation reveals that mainstream LLMs perform poorly on SV assessment tasks, largely due to a lack of assessment-specific knowledge.
To address the above challenges, we propose \appname, a novel framework that effectively leverages LLMs to perform SV assessment and provide supporting evidence.
\appname employs specialized LLM agents to process rich text content in \sirs and incorporate information about vulnerable projects.
\appname builds a dedicated assessment LLM by injecting assessment-specific knowledge through finetuning.
Specifically, we enable large-scale reasoning trajectory annotation using off-the-shelf LLMs and adopt a two-stage training paradigm, i.e., supervised instruction tuning to inject domain knowledge, followed by reinforcement learning to enhance the model’s intrinsic reasoning capability.
Evaluations on a newly collected \sir dataset demonstrate that \appname outperforms the best-performing baseline by 5.3\%-35.2\% across multiple evaluation metrics. 
Ablation studies validate the effectiveness of our design choices for both assessment-specific model training and SV information enrichment. 
Finally, a user study with security experts confirms that the evidence provided by \appname is useful and practical for real-world SV assessment.
\end{abstract}

\begin{CCSXML}
<ccs2012>
   <concept>
       <concept_id>10002978</concept_id>
       <concept_desc>Security and privacy</concept_desc>
       <concept_significance>500</concept_significance>
       </concept>
   <concept>
       <concept_id>10002978.10003022.10003023</concept_id>
       <concept_desc>Security and privacy~Software security engineering</concept_desc>
       <concept_significance>500</concept_significance>
       </concept>
 </ccs2012>
\end{CCSXML}

\ccsdesc[500]{Security and privacy}
\ccsdesc[500]{Security and privacy~Software security engineering}

\keywords{Software Security, Vulnerability Assessment, CVSS}

\maketitle

\vspace{-0.1cm}
\section{Introduction} \label{sec:introduction}
With the ever-growing volume of emerging software vulnerabilities (SVs), a critical challenge in SV management is to locate and prioritize the remediation of critical SVs~\cite{le2022survey}, i.e., those tend to be exploited and cause substantial damage (e.g., Log4Shell~\cite{CVE-2021-44228}).
Industry standards for SV handling processes (e.g., ISO/IEC 30111~\cite{iso_30111}) require software vendors to promptly address critical SVs. 
In addition, vendors must comply with security Service Level Agreements (SLAs)~\cite{SLA_phoenix, sla_hostedscan, cisa_bod}, which mandate the mitigation of critical SVs within a short timeframe (e.g., 15 days).

\begin{figure}[t]
    \centering
    \includegraphics[width=\linewidth]{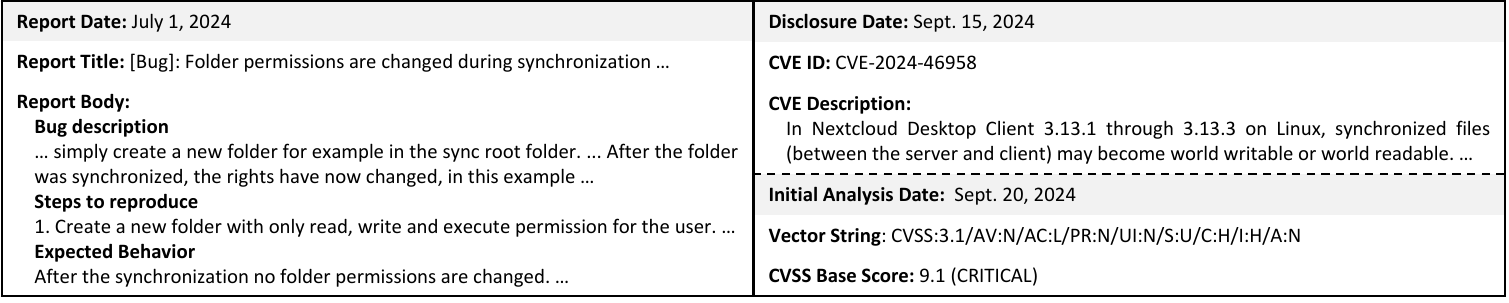}
    \vspace{-0.5 cm}
    \caption{CVE-2024-46958 and its corresponding report}
    \label{fig:motivating_example}
    \vspace{-0.3 cm}
\end{figure}

SV assessment, a critical phase in SV management life-cycle~\cite{le2022survey, foreman2019vulnerability}, corresponds to the process of prioritizing critical SVs.
Specifically, the \emph{assessment} phase characterizes SVs detected in the \emph{detection} phase to locate the ``hot spots", and supports to devise a prioritization plan for the \emph{remediation} phase.
Currently, Common Vulnerability Scoring System (CVSS) is the most widely used standard for SV assessment~\cite{CVSS_website}.
A common practice is to reference expert-curated CVSS metric values published in SV databases (e.g., NVD~\cite{NVD_website}) when devising the remediation schedule for newly disclosed SVs~\cite{imtiaz2022open, feutrill2018effect}.
However, manually assigning values to the CVSS metrics for the ever-growing number of SVs is both labor-intensive and time-consuming.
For example, the time lag between CVE disclosures and the availability of their CVSS metric values in the NVD has been increasing rapidly in recent years~\cite{feutrill2018effect, pan2024towards}.
This has motivated research on automated SV assessment~\cite{pan2024towards, le2022survey, le2021deepcva}.
In this work, we follow the majority studies~\cite{le2022survey} to use SV report (\sir) as the source to automate SV assessment (see Figure~\ref{fig:motivating_example} for an example).

Existing approaches typically train machine-learning (ML) or deep-learning (DL) models to classify \sirs into specific metric values~\cite{le2022survey}.
A significant drawback of existing assessment studies is their predominant focus on accuracy,
i.e., how accurate the results given by automated approaches are.
However, the performance of existing automated assessment approaches remains far from perfect~\cite{pan2024towards, le2022survey}.
Manual validation and curation remain inevitable in drawing the final assessment results, 
especially given that assessment outputs are referenced to determine remediation schedules~\cite{le2022survey}.
To effectively assist human analysts in performing efficient assessments, 
automated assessment approaches must not only provide their predictions but also present supporting evidence detailing their analysis in a manner that is easily comprehensible to humans.
Furthermore, existing approaches fail to fully exploit comprehensive SV information in two key aspects.
1)~They are unable to process information conveyed through rich-text content (e.g., screenshots and code snippets) embedded in \sirs. 
Such content is commonly used to illustrate attack steps and observed vulnerable behaviors, and is therefore essential for accurate SV assessment.
2)~They overlook information about the affected project (e.g., its functionality and usage scenarios), which provides context for ensuring correct assessments~\cite{allodi2018identifying}.

Given that Large Language Models (LLMs) have shown strong reasoning capabilities~\cite{wei2022chain} and have achieved promising results on SV-relevant tasks~\cite{zhou2024large}.
Applying LLMs to assess SVs and simultaneously present supporting evidence would be a promising direction.
In this work, we take the first step to comprehensively investigate the LLM performance in SV assessment.
Surprisingly, we find that applying LLMs to SV assessment is not straightforward. 
In fact, the performance of mainstream LLMs under various prompt settings falls significantly short of existing approaches by a wide margin.
Specifically, we find that while general LLMs possess necessary background knowledge to analyze a given \sir, they lack assessment-specific knowledge (e.g., assessment criteria), which leads to their struggling performance in SV assessment.

To this end, we propose \appname, a novel framework that effectively leverages LLMs to conduct SV assessment while presenting the supporting evidence.
\appname first utilizes three specialized LLM agents to analyze the code snippets and screenshots embedded in \sirs, and to retrieve information on vulnerable projects, respectively.

Then, we build a dedicated LLM by effectively injecting assessment-specific knowledge,
enabling it to autonomously analyze enriched SV information and make assessment decisions through structured reasoning.
In particular, we address two key challenges:
\ding{182}~Acquiring assessment-specific knowledge (e.g., assessment criteria, associations between metric values and specific SV features) at scale. 
Although expert-curated CVSS metric values are available in existing SV databases,
no dataset explains how these values are derived from SV features.
Producing such reasoning annotations with security experts would be prohibitively expensive.
To enable large-scale data construction using general-purpose LLMs, we ease the task of reasoning out the correct metric value into an \textit{open-book} problem.
We provide the LLM with the ground-truth metric value and the key attributes that distinguish it from alternative values, and prompt it to articulate the reasoning process based on the given SV information.
\ding{183}~
Instilling assessment-specific knowledge into the LLM and developing its intrinsic reasoning capability.
Compared with the \emph{open-book} annotation task, the \emph{closed-book} SV assessment task, i.e., autonomously analyzing an SV and inferring its metric values, is substantially more challenging.
Unlike prior efforts that transform general-purpose LLMs into domain specialists through a purely instruction-tuning “teacher–student” paradigm~\cite{Yu2025FineTuningLLMsCodeReview, mao2025towards}, our approach augments instruction tuning with reinforcement learning to cultivate the model’s inherent reasoning ability through online self-exploration~\cite{Guo2025DeepSeekR1}.
Moreover, we explicitly investigate whether learning to generate explicit reasoning trajectories improves assessment performance.

We evaluate \appname on a newly collected \sir dataset, which consists of \datasetsize \sirs from \datasetsizeproject projects. 
\appname is evaluated against a comprehensive collection of baselines, including six ML baselines, two DL baselines, and three advanced LLM baselines.
\appname consistently and significantly outperforms all baselines,
achieving improvements of 5.3\%-35.2\% over the best-performing baselines across various measures.
Ablation experiments verify the effectiveness of key designs including the two-stage finetuning and the processing of rich text contents.
We also conduct a user study, confirming that \appname can effectively help security analysts improve assessment efficiency.

In summary, this paper makes the following contributions:
\begin{itemize}[leftmargin=*]
    \item We present the first comprehensive investigation of SV assessment using LLMs and find that their performance is limited by the lack of domain-specific knowledge.
    We further propose a framework, \appname, to enable effective SV assessment using LLMs.
    
    \item \appname provides explicit reasoning to justify its assessment decisions and achieves strong performance through dedicated model construction, including large-scale dataset annotation, instruction tuning, and reinforcement learning. 
    To the best of our knowledge, this is the first application of reinforcement learning to SV assessment.
    
    \item \appname employs specialized LLM agents to effectively process rich text content in \sirs and integrate contextual knowledge about vulnerable projects.
    
    \item Extensive evaluations demonstrate that \appname significantly outperforms baselines, and a user study confirms its practical usefulness.
\end{itemize}

\section{Background and Related Work} \label{sec:related_work}
In this section, we introduce two aspects of the related work.

\noindent \textbf{SV Assessment with CVSS.}
CVSS is the \emph{de facto} standard for SV assessment~\cite{CVSS_website}.
CVSS is composed of three metric groups~\cite{cvss_v3_specification_document}:
1)~Base metrics, which capture the intrinsic properties of an SV;
2)~Temporal metrics, which reflect characteristics that change over the lifetime of an SV;
3)~Environmental metrics, which pertain to attributes of an SV that are specific to particular user environments.
Public assessments of SV severity (e.g., NVD~\cite{NVD_website} and vulnDB~\cite{vuldb}) focus exclusively on Base metrics, which represent the fundamental and unchanging characteristics of SVs over time and across different user environments.
In this work, we also focus on the Base metrics.
The Base group comprises eight specific metrics, as shown in Table~\ref{table:cvss_metrics}, which characterize SVs from two general perspectives: exploitability and impact.
After assigning values to each CVSS metric,
an overall severity score is calculated using standard equations~\cite{cvss_v3_specification_document}. 
This score can be further mapped to a qualitative rating (see Figure~\ref{fig:motivating_example} for an example).

\begin{table}[t]  %
  \centering
  \caption{Base metric group of CVSS}
  \vspace{-0.2cm}
  \scalebox{0.8}{
    \begin{tabular}{cll}
    \toprule
    \textbf{Category} & \textbf{Metric Names} & \textbf{Metric Values} \\
    \midrule
    \multirow{4}[2]{*}{Exploitability} & Attack Vector (AV) & Network, Adjacent, Local, Physical \\
          & Attack Complexity (AC) & Low, High \\
          & Privileges Required (PR) & None, Low, High \\
          & User Interaction (UI) & None, Required \\
    \midrule
    \multirow{3}[2]{*}{Impact} & Confidentiality Impact (C) & None, Low, High \\
          & Integrity Impact (I) & None, Low, High \\
          & Availability Impact (A) & None, Low, High \\
    \midrule
    -     & Scope (S) & Unchanged, Changed \\
    \midrule
    Overall  & Severity & Low, Medium, High, Critical \\
    \bottomrule
    \end{tabular}}
    \vspace{-0.3cm}
  \label{table:cvss_metrics}%
\end{table}%

\noindent \textbf{Automated SV Assessment.}
Le~\emph{et al.}~\cite{le2022survey} conducted a comprehensive literature review summarizing research on data-driven SV assessment and prioritization.
According to their survey, CVSS is the primary standard and SV reports are the mostly used source to develop automated SV assessment approaches.
Existing studies generally train ML or DL classifiers to predict the overall severity metric~\cite{han2017learning, li2023prediction, kudjo:inverse_gravity_moment:2019} or the entire group of CVSS metrics~\cite{le2021deepcva, le2022use, le2024mitigating, pan2024towards, spanos2018multi}.
Some recent and representative studies are:
Le~\emph{et al.}~\cite{le2021deepcva} suggest that the prediction of specific CVSS metrics may involve shared features and, therefore, 
propose a model with a unified encoder for the prediction of all CVSS metrics.
They demonstrate that adopting multitask learning outperforms developing separate models for each metric.
Pan~\emph{et al.}~\cite{pan2024towards} point out that specific CVSS metrics are strongly associated and further propose a prompt-tuning based approach (namely \proEVA) to exploit such relations for CVSS metrics prediction.

Different from existing works, we are the first to systematically investigate the application of LLMs for SV assessment. 
Additionally, we highlight the importance of providing evidence to justify predicted assessment results, a critical aspect overlooked by prior research.

\section{Motivation and Preliminaries} \label{sec:motivation}
In this section, we outline the challenges faced by the current SV assessment works.
Then, we conduct a comprehensive investigation into the application of LLMs in SV assessment.

\subsection{Challenges for SV Assessment} \label{subsec:motivation_challenge}
Existing automated assessment approaches
train ML or DL models
to classify \sirs into metric values~\cite{le2022survey, pan2024towards}, which suffer from the following three limitations:

\noindent \textbf{Impracticality of assessment results without supporting evidence.}
Existing approaches~\cite{le2021deepcva, pan2024towards, le2022survey} 
have exclusively focused on the assessment accuracy, 
i.e., how accurately the algorithms assign correct CVSS metric values to a given SV.
However, manual validation and curation remain inevitable, 
as the reliability of assessment results is critical for guiding remediation schedules~\cite{le2022survey, dissanayake2022empirical, foreman2019vulnerability} and yet the performance of automated SV assessment still falls short of perfection.
To assist security analysts in validating the automated assessment results, evidence supporting the model decision is essential.
Unfortunately, this has been overlooked by existing studies.

We propose to provide supporting evidence (e.g., extracted keywords and phrases) from the given SV information, together with the necessary analysis to justify the model’s predictions.
This evidence highlights the key information relevant to SV assessment, enabling analysts to quickly develop a comprehensive understanding of the vulnerability and to efficiently validate the automated assessment results.
Figure~\ref{fig:approach_overview} illustrates an example, showing how the inclusion of detailed evidence and analysis (under the \verb|[Clues]| and \verb|[Reasoning]| tags) facilitates more effective manual validation and curation.
LLMs have the potential to provide assessment results with evidence given their strong reasoning capability~\cite{wei2022chain, bang2023multitask}.
However, as we show in Section~\ref{subsec:motivation_llm_reasoning}, it is not trivial to leverage LLMs to conduct SV assessment.

\noindent \textbf{Ineffectiveness in processing \sirs.}
\ding{182}~\sirs are not plain texts but embedded with multiple rich text content, e.g., screenshots and code snippets.
24.0\% of our collected \datasetsize \sirs (see Section~\ref{subsec:data_collection}) contain at least one screenshot, and 10.1\% contain over three screenshots.
For code snippets, 69.2\% of our collected \sirs contain at least one, and 16.8\% contain over five.
Screenshots are commonly used to present attack steps and observed vulnerable behaviors.
For example, the report~\cite{glazedlists_issue_709} of CVE-2023-31890 (XML de-serialization without proper checks) uses a screenshot to show that the provided attack payload can successfully exploit SV to launch the calculator application.
This screenshot helps to confirm that exploiting SV can result in arbitrary code execution, 
which is essential to determine the value of impact metrics.
Code snippets embedded in \sirs could be Proof-of-Concept (PoC) snippets, Address Sanitizer (ASAN) reports, etc.
However, existing approaches fail to process the information conveyed by rich text content.
A common practice is to replace such content with placeholder tags~\cite{pan2024towards}, for example, substituting screenshot URLs with a generic \verb|IMGTAG|.
This abstraction discards critical details, such as attack steps or vulnerable behaviors, embedded in the rich content, which can in turn lead to incomplete understanding and inaccurate SV assessment.
\ding{183}~The code snippets embedded in \sirs are often lengthy and noisy.
In our collected dataset, 22.1\% of \sirs contain embedded code snippets exceeding 2,000 characters (approximately 500 tokens).
These long snippets typically have low information density, yet they consume a substantial portion of the LLMs’ context window and distract the model from other critical information, 
thereby increasing the likelihood of incomplete reasoning and inaccurate SV assessment results.

To address these challenges, 
we propose first leveraging specialized LLM agents to analyze the screenshots and code snippets embedded in \sirs, and then replacing these rich textual contents with concise analytical summaries before performing the final assessment.
This pipeline decomposes the complex analysis of \sirs and reduces the context length required at each analysis stage.

\noindent \textbf{Lack of information regarding the vulnerable project.}
Existing approaches rely solely on the SV report itself as input, overlooking contextual information about the vulnerable project (e.g., its functionality and usage scenarios), which is often helpful for accurate SV assessment.

For the SV shown in Figure~\ref{fig:motivating_example}, the report only states that the vulnerability involves file system synchronization. 
With this limited information, \emph{Local}, \emph{Adjacent}, and \emph{Network} are all plausible values for the \emph{Attack Vector} metric. 
However, knowing that the affected project is \texttt{nextcloud/desktop}, a cloud storage and file-sharing application, makes it clear that \emph{Network} is the correct value.

Vulnerable project information is recognized by established standards (e.g., NIST 800-30~\cite{stoneburner2002risk} and Common Criteria~\cite{infrastructure2002common}) and prior studies~\cite{allodi2018identifying} as essential context for ensuring accurate SV assessment.

We propose enriching \sirs with information about the vulnerable projects to enable a more comprehensive assessment.

\subsection{General LLM for SV Assessment} \label{subsec:motivation_llm_reasoning}

\begin{figure}[t]
    \centering
    \includegraphics[width=\linewidth]{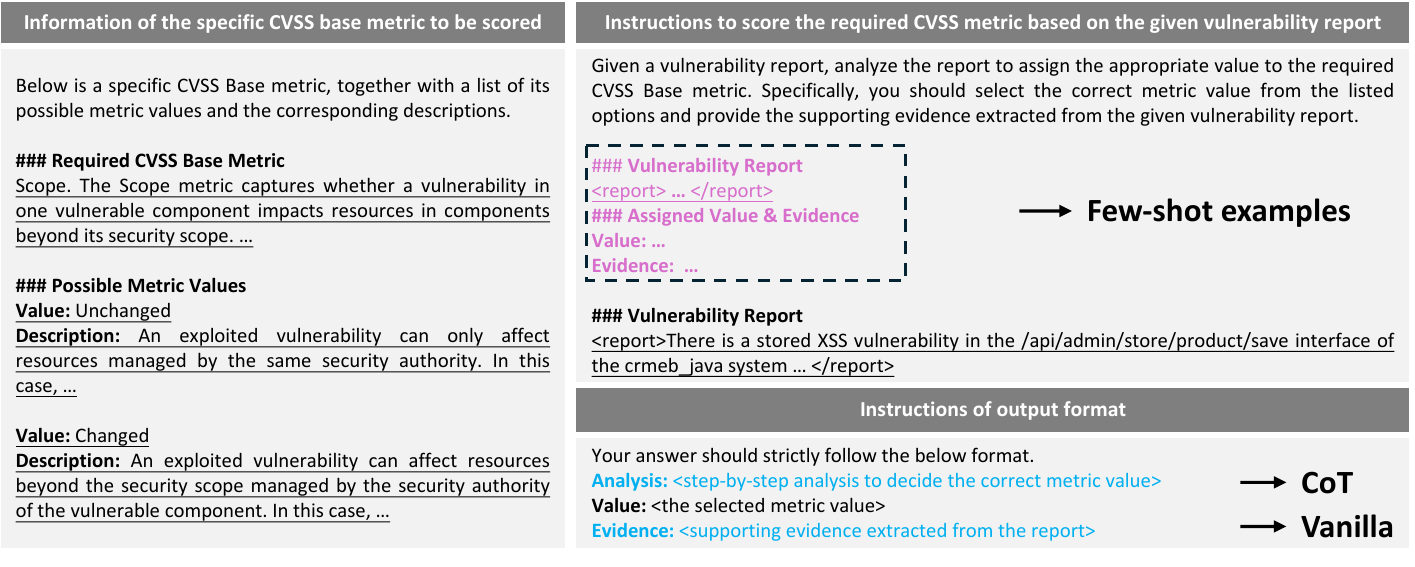}
    \vspace{-0.6 cm}
    \caption{Prompt templates used in the preliminary study}
    \label{fig:prompt_preliminary}
    \vspace{-0.6 cm}
\end{figure}

A straightforward approach to obtaining assessment results, supported by evidence, is to prompt the LLM to reason over the input SV information and draw a conclusion.
However, after comprehensive evaluations, we find that off-the-shelf LLMs struggle to make correct assessments. 
Specifically, we evaluate the performance of three advanced LLMs (i.e., Llama-3.3-70b~\cite{llama_3.3}, DeepSeek-v3~\cite{liu2024deepseek}, 
and GPT-4.1~\cite{gpt_4.1}) 
under the following four prompt methods:
\ding{182}~zero-shot vanilla prompting,
\ding{183}~zero-shot Chain-of-Thought (CoT) prompting~\cite{wei2022chain},
\ding{184}~few-shot vanilla prompting,
\ding{185}~few-shot CoT prompting.
In total, our evaluation considers 12 distinct combinations.
Refer to Section~\ref{sec:exp_setup} for detailed evaluation settings (e.g., evaluation set and metrics).

Figure~\ref{fig:prompt_preliminary} shows the prompt templates used in our evaluation.
The templates are parameterized with input-specific content, indicated by underlined text. 
Each prompt consists of three segments:
\begin{itemize}[leftmargin=*]
    \item \textbf{Information of the specific CVSS metric to be scored.} 
    We prompt the LLM to score one CVSS metric at a time, since each metric characterizes an independent aspect of an SV. 
    We supply LLM with the metric introduction and candidate metric values along with their descriptions extracted from the CVSS specification~\cite{cvss_v3_specification_document}.
    In preliminary experiments, omitting this detailed metric information led to substantially worse performance.
    \item \textbf{Instructions to score the required CVSS metric.}
    We instruct the LLM to first analyze the given \sir, then select the correct metric value from the listed options, and provide supporting evidence.
    For few-shot prompt setting, to the best of our knowledge, the only public source that provides evidence or explanations for the assigned CVSS metric value is the official CVSS document~\cite{CVSS_examples}.
    The document presents 31 examples with expert-curated evidence to demonstrate how to apply CVSS to score specific SVs.
    We select SVs from these examples to form the few-shot prompt.
    Specifically, we first apply a hybrid retriever (see Section~\ref{subsec:approach_retrieve}) to rank these examples based on their similarities to the given SVR and then select one for each candidate metric value.
    \item \textbf{Instructions of output format.}
    For vanilla setting, we instruct the LLM to first give the verdict and then further provide evidence.
    For CoT setting, we instruct the LLM to first perform a step-by-step reasoning before making the final verdict.
    The reasoning process is regarded as evidence justifying the model decision.
\end{itemize}

Table~\ref{table:preliminary} presents the average performance on eight CVSS metrics for each LLM-prompting combination.
The results show that the performances of all three general LLMs under various prompt settings are much worse than existing learning-based approachs (i.e., \proEVA~\cite{pan2024towards}).
Moreover, we observe that neither stronger LLMs nor CoT prompting necessarily lead to improved assessment performance.
Under zero-shot settings, Llama-3.3-70B achieves performance comparable to DeepSeek-V3 and GPT-4.1, 
despite the latter demonstrating substantially stronger performance on general benchmarks~\cite{llm_leaderboard}.
Under both zero- and few-shot settings, 
CoT prompting yields performance that is comparable to, or slightly worse than, that achieved with vanilla prompting across all three LLMs.
In contrast, providing LLMs with additional domain knowledge through few-shot examples consistently and substantially improves the performance of all three models compared to their zero-shot counterparts.

To further understand the role of domain-specific knowledge in SV assessment, we manually examine cases that are correctly assessed by \proEVA but misclassified by LLMs. 
We identify two primary reasons for these errors:

\begin{wraptable}{r}{0.5\textwidth}
  \centering
  \caption{The average performance on eight CVSS metrics for three LLMs under four prompt settings}
  \vspace{-0.1cm}
  \scalebox{0.85}{
    \begin{tabular}{clcc}
    \toprule
    \textbf{Prompt} & \textbf{LLM} & \textbf{F1} & \textbf{MCC} \\
    \midrule
    \multirow{3}[2]{*}{\textbf{\tabincell{c}{Zero-Shot\\Vanilla}}} & \multicolumn{1}{l}{\textbf{Llama}} & 0.747  & 0.330  \\
          & \multicolumn{1}{l}{\textbf{DeepSeek}} & 0.746  & 0.338  \\
          & \multicolumn{1}{l}{\textbf{GPT}} & 0.736  & 0.358  \\
    \midrule
    \multirow{3}[2]{*}{\textbf{\tabincell{c}{Zero-Shot\\CoT}}} & \multicolumn{1}{l}{\textbf{Llama}} & 0.735  & 0.314  \\
          & \multicolumn{1}{l}{\textbf{DeepSeek}} & 0.726  & 0.318  \\
          & \multicolumn{1}{l}{\textbf{GPT}} & 0.740  & 0.362  \\
    \midrule
    \multirow{3}[2]{*}{\textbf{\tabincell{c}{Few-Shot\\Vanilla}}} & \multicolumn{1}{l}{\textbf{Llama}} & 0.778  & 0.402  \\
          & \multicolumn{1}{l}{\textbf{DeepSeek}} & 0.783  & 0.429  \\
          & \multicolumn{1}{l}{\textbf{GPT}} & 0.781  & 0.485  \\
    \midrule
    \multirow{3}[2]{*}{\textbf{\tabincell{c}{Few-Shot\\CoT}}} & \multicolumn{1}{l}{\textbf{Llama}} & 0.745  & 0.346  \\
          & \multicolumn{1}{l}{\textbf{DeepSeek}} & 0.758  & 0.399  \\
          & \multicolumn{1}{l}{\textbf{GPT}} & 0.765  & 0.456  \\
    \midrule
    \multicolumn{2}{c}{\textbf{proEVA}} & 0.831  & 0.544  \\
    \bottomrule
    \end{tabular}}
    \label{table:preliminary}
    \vspace{-0.1cm}
\end{wraptable}

\noindent \textbf{SV reports lack sufficient information to determine the metric value.}
In such cases, the assessment process is largely experience- or knowledge-based~\cite{cvss_v3_examples}, 
with historical SVs with similar characteristics (e.g., root cause) serving as important information cues.
A typical case is memory-related SVs (e.g., use-after-free, heap buffer overflow).
Reports of this category of SVs often only present an ASAN output showing that the SV can be successfully triggered with the provided PoC, without detailing its potential impact (e.g., CVE-2024-6064~\cite{gpac_issue_2974}).
As a result, determining the values of the impact metrics for such SVs heavily relies on referencing similar historical cases.
For example, 
similar to CVE-2024-6064, CVE-2022-27147 (a historical SV in the training set of \proEVA) is also a use-after-free SV in the GPAC project, which shares the same values for all three impact metrics.

\noindent \textbf{LLMs fail to identify implicit clues in the report signaling specific metric values or LLMs misinterpret the assessment criteria.}
A typical failure case involves determining the value of the \emph{Scope} metric for XSS SVs.
For stored XSS SVs, the \emph{Scope} metric is always \verb|Changed|~\cite{CVSS_examples}, because the vulnerable component is the web server (where the malicious script is injected), whereas the impacted component is the victim’s browser (where the script is executed).
However, the best accuracy among all three LLMs under zero-shot vanilla/CoT promptings is only 15.3\% (11 out of 72).
Taking CVE-2023-1609 as an example, 
the corresponding report explicitly mentions that it is a stored XSS SV (see Figure~\ref{fig:prompt_preliminary}). 
LLM correctly analyzes the SV but incorrectly assigns the value \verb|Unchanged| by stating that \emph{``Both the vulnerable component (the API endpoint that stores the payload into the database) and the impacted component (the front-end page that renders the payload) are managed by the same security authority (the crmeb\_java application)"}.
We observe that by providing similar examples (i.e., few-shot prompting), 
the accuracy of LLMs is largely improved (from 11 to 59), though still lower than \proEVA (correctly predicts all 72 SVs).

We further observe that given the metric value (\verb|Changed|), LLM can identify Stored XSS as one supporting evidence. 
Still taking CVE-2023-1609 as an example, LLM states that \emph{``This indicates that an attacker can inject malicious code into the system, which can then be executed by other users, affecting resources beyond the initial security scope"}.
This suggests that, although LLMs struggle to accurately reason out the appropriate metric value, they possess the necessary background knowledge to analyze SVs and identify relevant evidence for a given metric value.
Identifying supporting evidence for a provided answer (i.e., a \emph{open-book} problem) is considered easier and more straightforward than selecting the correct answer through reasoning (i.e., a \emph{closed-book} problem).

To summarize, our preliminary findings show that 
\ding{182}~it is not trivial to leverage general LLMs to conduct SV assessment.
The primary reason is their lack of knowledge specific to SV assessment (e.g., historical SV information and assessment criteria);
\ding{183}~though we show that supplying LLM with the assessment knowledge (i.e., the metric introduction and the demonstrating examples) extracted from the CVSS official document improves the performance, the improvement is still limited.
These findings motivate us to finetune LLMs to effectively inject domain-specific knowledge rather than directly prompting off-the-shelf LLMs to conduct assessment.

\section{Approach} \label{sec:approach}
\begin{figure*}[t]
    \centering
    \includegraphics[width=\linewidth]{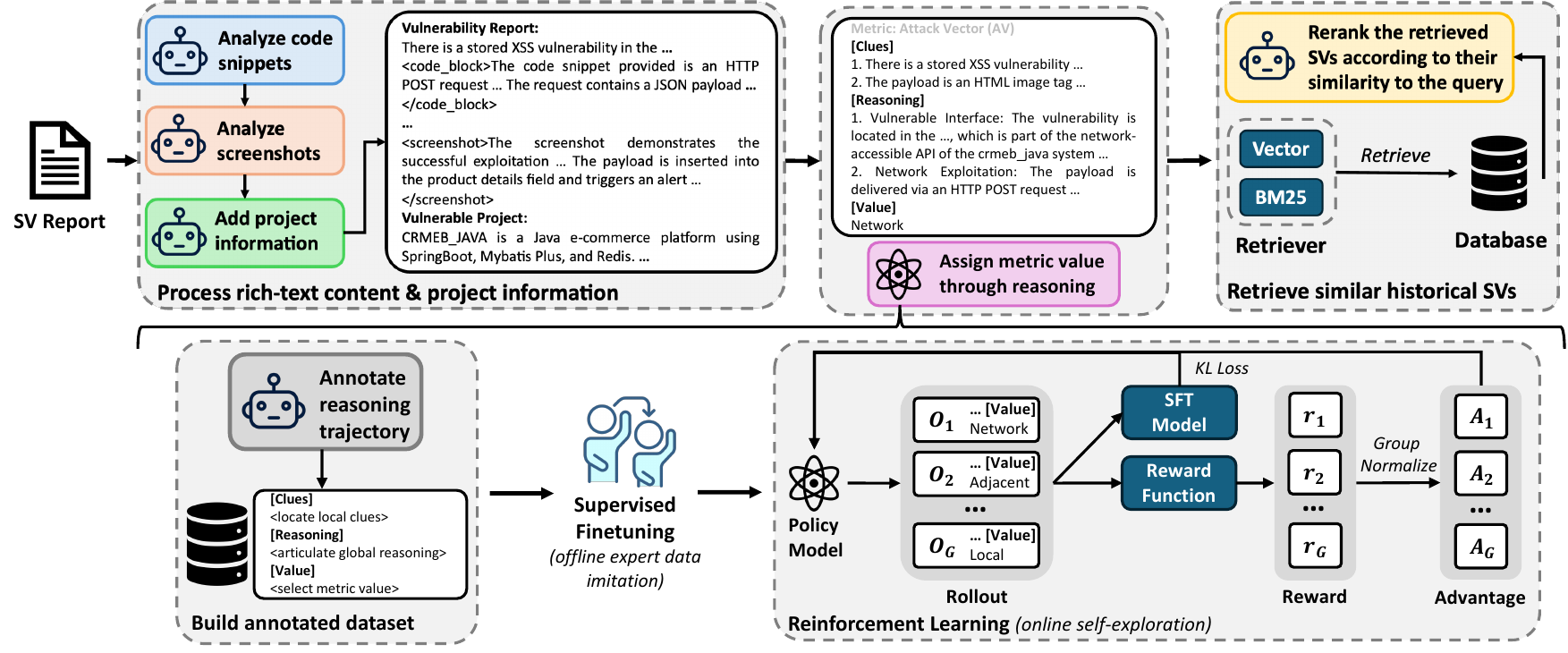}
    \vspace{-0.5 cm}
    \caption{Overview of \appname}
    \label{fig:approach_overview}
    \vspace{-0.5 cm}
\end{figure*}

\vspace{-0.1cm}
In this section, we introduce our proposed approach, namely \appname.
Figure~\ref{fig:approach_overview} presents the overview of \appname.
Before applying \appname to assess future SVs, 
we first construct a dataset by employing an LLM to annotate reasoning trajectories for disclosed SVs using analyst-curated assessment results.
This annotated dataset is then used to finetune a dedicated assessment LLM and to build the database for retrieving similar historical SVs. 
Specifically, we adopt a two-stage finetuning paradigm that combines supervised finetuning with reinforcement learning.
During inference, given an SV report (\sir), \appname first employs three specialized LLM agents to analyze the embedded code snippets and screenshots, as well as to incorporate information about the vulnerable project.
The resulting enriched SV information is fed into the finetuned assessment LLM to reason out the appropriate metric value.
The generated reasoning trajectory is subsequently transformed into an index to retrieve similar historical SVs, which are provided as supplementary supporting evidence.

\subsection{Processing Rich Text Content and Project Information} \label{subsec:approach_richtext}

We build specialized LLM agents to analyze the rich text content in the \sir and summarize the information of the vulnerable project.
The details are as follows:

\noindent \textbf{Analyze code snippets.}
We first use regular expressions to extract embedded code snippets.
Since replacing the original code snippet with a descriptive summary derived from the analysis may result in information loss, 
we limit our processing to exceptionally lengthy code snippets (i.e., those exceeding 500 characters) that exhibit low information density and are likely to distract LLMs from other crucial information during SV assessment. 
We analyze each extracted code snippet one at a time, following their order in the report.
Specifically, considering that necessary context helps to enhance the understanding of information conveyed by the code snippet,
we supply the preceding context from the \sir to the LLM and prompt it to describe the content of the code snippet under analysis.

The prompt template (refer to online appendix~\cite{replication_submit}) is composed of the following three segments:

\ding{182}~Instructions to analyze the code snippet. We specify that the code snippet is presented in a \sir and the analysis should focus on SV-relevant aspects.
\ding{183}~Code snippet to be analyzed along with its preceding context. 
\ding{184}~Instructions of output format. We instruct the LLM to conduct a two-step CoT-style analysis, i.e., a comprehensive look through followed by a brief summary.

\noindent \textbf{Analyze screenshots.}
We first use regular expressions to extract URLs of screenshots embedded in \sirs.
Then, we download the screenshot and encode it into a Base64~\cite{base64} string, which can be processed by LLM services.
We use a multimodal LLM, which is capable of conducting image comprehension, to perform the analysis.
Similar to the analysis of code snippets, we analyze each extracted screenshot sequentially following their order in the report, and provide the preceding context to help LLM better interpret the conveyed information.

The prompt template (refer to online appendix~\cite{replication_submit}) shares a similar design with the one used for code snippet analysis.

\noindent \textbf{Add project information.}
We first crawl the project introduction from its homepage and then prompt the LLM to give a summarization emphasizing its functionalities and applications.

\subsection{Building Dedicated Assessment LLM through Finetuning} \label{subsec:approach_finetune}
To turn a general-purpose LLM into an SV assessment expert, we first construct large-scale reasoning trajectories that illustrate how expert-curated CVSS metric values are derived for historical SVs. 
We then apply supervised instruction tuning (SFT) to inject assessment-specific knowledge, followed by reinforcement learning (RL) to further enhance the model's reasoning capability.

\subsubsection{Annotating Reasoning Trajectories} \label{subsubsec:approach_finetune_dataset}

\begin{figure}[t]
  \centering
  \begin{subfigure}[t]{0.49\linewidth}
    \centering
    \includegraphics[width=\linewidth]{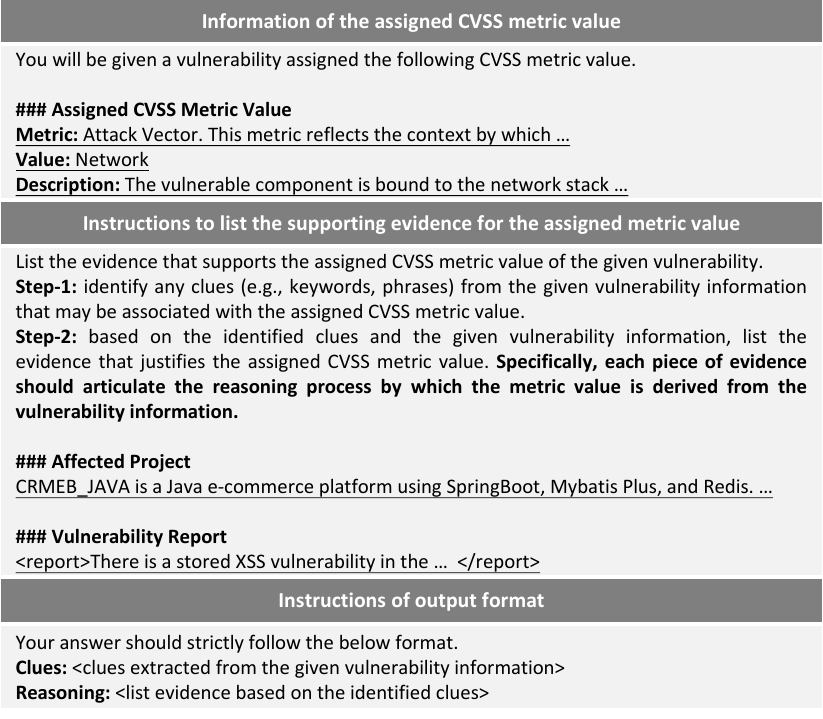}
    \caption{\todo{Prompt for annotating reasoning trajectories}}
    \label{fig:evidence_prompt}
  \end{subfigure}\hfill
  \begin{subfigure}[t]{0.49\linewidth}
    \centering
    \includegraphics[width=\linewidth]{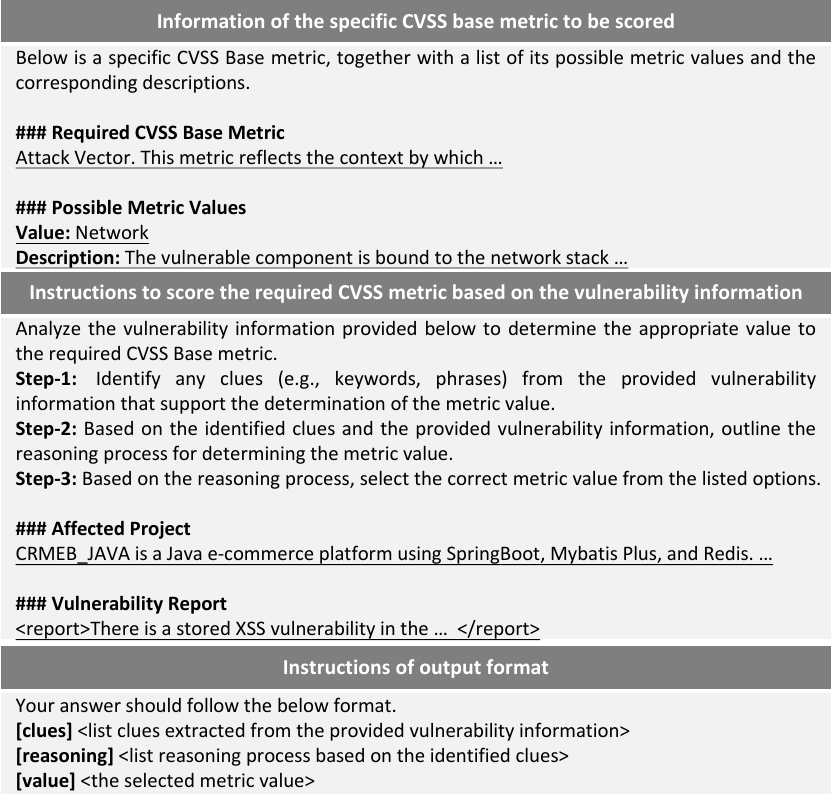}
    \caption{\todo{Prompt for finetuning LLM for SV assessment}}
    \label{fig:finetune_prompt}
  \end{subfigure}
  \vspace{-0.1cm}
  \caption{Prompt templates used in data annotation and finetuning}
  \vspace{-0.3cm}
\end{figure}

Although existing SV databases provide expert-curated CVSS metric values, there is currently no annotated dataset that captures the assessment process, specifically, how these metric values are derived from SV features. 
Constructing such annotations with the involvement of security experts would incur substantial cost. 
Leveraging LLMs for data synthesis has shown promise across various domains~\cite{Yu2025FineTuningLLMsCodeReview}.
However, SV assessment requires substantial domain expertise, and our preliminary study (see Section~\ref{subsec:motivation_llm_reasoning}) indicates that general-purpose LLMs struggle to produce accurate assessment results.
To facilitate large-scale data annotation, we therefore reformulate the task of deriving the correct metric value as an \textit{open-book} problem. 
Under this formulation, the LLM is provided with the ground-truth metric value along with key distinguishing attributes, and is prompted to articulate the reasoning trajectory based on the SV information.
The detailed prompt template is shown in Figure~\ref{fig:evidence_prompt}.
The first segment of the prompt supplies information about the assigned CVSS metric value, including a brief introduction to the metric and the key features that distinguish the assigned value from other candidates, as extracted from the official CVSS specification~\cite{cvss_v3_specification_document}.

\camera{To ensure the quality of the open-book annotations, we develop the prompt through an iterative refinement process. Specifically, we conduct pilot experiments on a randomly sampled subset, in which the first two authors independently evaluate the generated evidence in terms of faithfulness and completeness. Based on the identified issues, they discuss potential improvements, refine the prompt, and repeat this process over multiple rounds.}
The resulting structured prompt is designed based on Chain-of-Thought (CoT) techniques~\cite{wei2022chain, sun2023text} to improve both the comprehensiveness and faithfulness of the annotated reasoning trajectories. It decomposes the reasoning process into two steps that mirror the analysis workflow of a security analyst. As illustrated in the second segment of the prompt in Figure~\ref{fig:evidence_prompt}, we first instruct the LLM to examine the given SV information and identify relevant \emph{clues}, i.e., local factual features such as keywords and phrases that indicate the assigned metric value. After locating these clues, the LLM is guided to synthesize the local evidence into a global understanding and articulate the reasoning process through which the metric value is derived.
This locate-and-analysis process offers the following benefits~\cite{sun2023text}:
1) decomposing the analysis allows LLMs to focus on individual sub-goals, leading to more comprehensive evidence localization and deeper reasoning;
2) initially collecting surface-level keywords and phrases ensures that the raw evidence, used as the starting point for subsequent reasoning, faithfully originates from the given SV information.

We apply this approach to annotate reasoning trajectories for 51,568 metric-level data points, corresponding to the eight CVSS metrics for each of the 6,446 SVRs collected in Section~\ref{subsec:data_collection}.
This effort fills a significant data gap in demonstrating the SV assessment process.
To validate the quality of the annotated data, we randomly sample 400 data points and engage two domain experts to interdependently evaluate each samples against two criteria: 
factuality (i.e., whether the evidence is grounded in the given SV information) and relevance (i.e., whether the evidence directly supports the assigned metric value).
The validation process required a total of 56 person-hours.
The two experts achieved a Cohen’s Kappa score of 0.84, indicating substantial inter-rater agreement.
Overall, 391 data points (97.8\%) satisfy both quality criteria, confirming the effectiveness of our proposed automated annotation approach.
In addition, we indirectly evaluate the quality of reasoning trajectories produced by the finetuned model through a userstudy (see Section~\ref{subsec:RQ4}).

\subsubsection{Finetuning LLM for SV Assessment} \label{subsubsec:approach_finetune_rl}
We further leverage the annotated reasoning trajectories of historical SVs to train a dedicated LLM for SV assessment, \todoa{which is substantially smaller than the model used for data annotation.}
Specifically, the model input is the enriched SV information from the previous step and the output is a reasoning process that leads to a determined metric value.

As demonstrated in our preliminary study, general-purpose LLMs lack assessment-specific knowledge. 
We therefore first apply supervised instruction tuning (SFT)~\cite{peng2023instruction} to inject domain knowledge at scale. The instruction templates used for fine-tuning are shown in Figure~\ref{fig:finetune_prompt}. 
Specifically, we instruct the LLM to reason through three steps that correspond to the \emph{locate-and-analysis} Chain-of-Thought used during reasoning trajectory annotation:
1)~locating key clues,
2)~outlining the reasoning process, and
3)~selecting the correct metric value.
The LLM is fine-tuned to imitate the annotated reasoning trajectories by optimizing the causal language modeling loss~\cite{radford2018improving}.

Although the instruction-tuned LLM can acquire some level of assessment reasoning ability by imitating expert trajectories, it still suffers from two inherent limitations.
\ding{182}~There exists a discrepancy between the \emph{open-book} setting used during data annotation and the \emph{closed-book} reasoning capability required at inference time.
\ding{183}~SFT is known to exhibit limited generalization ability, i.e., tuned models often struggle to transfer leaned reasoning patterns across different settings~\cite{chusft, zhang2021bridging}. 
This limitation is particularly pronounced in our SV assessment task due to the discrepancy between the annotation and inference settings.
Thus, in contrast to prior efforts that transform general LLMs into domain specialists using a single teacher–student instruction-tuning paradigm~\cite{Yu2025FineTuningLLMsCodeReview, mao2025towards}, 
we extend conventional SFT with reinforcement learning (RL).
This design enables the model to develop intrinsic reasoning capabilities through online self-exploration~\cite{Shao2024DeepSeekMath}.

Considering that the SV assessment task can be automatically and reliably verified, by comparing the model’s predicted severity level with the ground truth, we adopt the the Group-relative Policy Optimization (GRPO) algorithm~\cite{Shao2024DeepSeekMath} to further refine the instruction-tuned model.
Different from general RL algorithms like Proximal Policy Optimization (PPO) that requires training a separate value model~\cite{schulman2017proximal}, GRPO estimates advantages in a group-relative manner and can thus take advantage of verifiable outcome to reduce complexity in reward computing and assign consistent reward signals.
Specifically, we define the reward function as follows:

\begin{equation}
\mathcal{R}(o) =
\begin{cases}
0, & \text{if the output format is invalid}, \\[4pt]
0.1,  & \text{if } \mathrm{value}_{\text{pred}}^{m} \text{ is not a valid metric value}, \\[4pt]
1.5 - \dfrac{\mathrm{dis}(\mathrm{value}_{\text{pred}}^{m}, \mathrm{value}_{\text{gt}}^{m})}
{\mathrm{max\_dis}^{m}}, & \text{otherwise.}
\end{cases}
\label{eq:reward_function}
\end{equation}

\camera{Here}, 
$\mathrm{dis}(\mathrm{value}_{\text{pred}}^{m}, \mathrm{value}_{\text{gt}}^{m})$ measures the distance between the predicted and ground-truth metric values,
and $\mathrm{max\_dis}^{m}$ represents the maximum possible distance for the CVSS metric $m$.
The distance is defined as the absolute difference between the indices of the predicted and ground-truth values for a given CVSS metric, i.e., $\mathrm{dis}(\mathrm{value}_{\text{pred}}^{m}, \mathrm{value}_{\text{gt}}^{m}) = \left | \operatorname{idx}(\mathrm{value}_{\text{pred}}^{m}) -  \operatorname{idx}(\mathrm{value}_{\text{gt}}^{m}) \right |$. For example, for the \emph{Attack Vector} metric, its four candidate values (i.e., \emph{Network}, \emph{Adjacent}, \emph{Local}, and \emph{Physical}) are indexed as 0, 1, 2, and 3, respectively.
Thus, the distance between \emph{Network} (0) and \emph{Local} (2) is 2, while $\mathrm{max\_dis}^{m}$ is 3, corresponding to the distance between \emph{Network} (0) and \emph{Physical} (3) (see Table~\ref{table:cvss_metrics}).
Intuitively, this distance reflects the semantic gap between the predicted and ground-truth values, with larger distances indicating greater mismatches. This reward design encourages the LLM to generate outputs that are both syntactically valid and semantically close to the ground truth. Improperly formatted or invalid outputs are penalized, while predictions closer to the correct value receive higher rewards.

Given a training set $\mathcal{D}$ where each instance consists of
1)~$\mathrm{SVR}$, enriched SV information described in Section~\ref{subsec:approach_richtext};
2)~$m$, one of the eight CVSS metrics;
3)~$\mathrm{value_{gt}}$, the ground-truth metric value.
The input prompt is formatted using the template shown in Figure~\ref{fig:finetune_prompt}, denoted as $q = \mathrm{prompt} \left (\mathrm{SVR}, m \right )$.
For each sampled instance, 
the policy LLM $\pi_\theta$ tries to assign the appropriate metric value through reasoning multiple times, i.e., generating $G$ (the group size) outputs given the prompt $q$.
Each output $o_i$ is scored using the reward function defined above and normalized within the group to get the corresponding group-relative advantage $A_i$.
Finally, the policy LLM is optimized by maximizing the following GRPO objective (see Figure~\ref{fig:approach_overview}):

\begin{equation}
\begin{split}
\mathcal{J}(\theta) = 
&\; \mathbb{E} \Bigg[
\frac{1}{G} \sum_{i=1}^{G}
\Big(
\min\!\Big(
\frac{\pi_\theta(o_i \mid q)}{\pi_{\theta_{\text{old}}}(o_i \mid q)} A_i,\;
\text{clip}\!\Big(
\frac{\pi_\theta(o_i \mid q)}{\pi_{\theta_{\text{old}}}(o_i \mid q)},
1-\epsilon,\;1+\epsilon
\Big) A_i
\Big)
\\[4pt]
&\quad
- \beta D_{\mathrm{KL}}\!\big(\pi_\theta \,\|\, \pi_{\text{ref}}\big)
\Big)
\Bigg],
\\[6pt]
\text{where } 
&\; (\mathrm{SVR}, m, \mathrm{value_{gt}}) \sim \mathcal{D},
\ q = \mathrm{prompt}(\mathrm{SVR}, m),
\ \{o_i\}_{i=1}^{G} \sim \pi_{\theta_{\text{old}}}(\cdot \mid q).
\end{split}
\label{eq:grpo_function}
\end{equation}

Here, $\epsilon$ and $\beta$ are hyperparameters, $\pi_{\theta_{\text{old}}}$ and $\pi_{\theta_{\text{ref}}}$ denote the old and reference policies, respectively.
The ratio $\frac{\pi_\theta(o_i \mid q)}{\pi_{\theta_{\text{old}}}(o_i \mid q)}$ measures the relative likelihood of generating the output $o_i$ under the new policy $\pi_\theta$ versus the old policy $\pi_{\theta_{\text{old}}}$, 
and is used to correct the advantage $A_i$ for distributional shift.
To stabilize training, this ratio is clipped to $[1-\epsilon, 1+\epsilon]$, and the objective for each token is defined as the minimum of the clipped and unclipped terms.
A KL-divergence penalty $\beta D_{\mathrm{KL}}$ regularizes the policy update to limit deviation from the reference policy.
The token-level objectives are averaged over all tokens and outputs in the batch to form the final loss.

\vspace{-0.1cm}
\subsection{Retrieving Similar Historical SVs} \label{subsec:approach_retrieve}
After the fine-tuned LLM generates the assessment prediction, we retrieve similar historical SVs to enrich the supporting evidence. The historical SV retrieval module does not participate in generating the prediction and therefore does not affect the assessment results. Instead, it uses the model-generated reasoning trajectory to retrieve similar historical SVs as post-hoc evidence. Presenting analysts with historical SVs that share similar characteristics and are assigned the same metric value helps strengthen the credibility and interpretability of the assessment result, particularly when \sirs provide limited information and the assessment relies heavily on expert experience (see Section~\ref{subsec:motivation_llm_reasoning}). The retrieval process is described below:

\noindent \textbf{Index/Retrieval proxy.}
A straightforward way is to use the raw \sir to retrieve similar historical SVs.
However, the raw \sirs are often noisy (e.g., containing code snippets, URLs), 
unstructured,
and include SV-irrelevant information (e.g., replication environments).
Moreover, each CVSS metric captures a distinct aspect of an SV, whereas raw \sirs are not tailored to any specific metric,
failing to achieve a precise metric-specific retrieval.

We propose to utilize the evidence collected from the raw \sir and further organized by the LLM (i.e., the reasoning trajectory) as the retrieval proxy.

\noindent \textbf{Retriever.}
We conduct a hybrid retrieval~\cite{luan2021sparse} by assembling a \emph{sparse} retriever and a \emph{dense} retriever.
The sparse and dense retriever individually perform a search and return top-\emph{k} historical SVs based on keywords and semantics, respectively. 
Their outputs are assembled to get the final retrievals.

\noindent \textbf{LLM re-ranker.}
After using the built hybrid retriever to narrow down the pool of candidate historical SVs, we leverage an LLM as a reranker to further assess the relevance of the retrieved SVs to the query SV.
Notably, we batch the retrieved historical SVs and query the LLM to rate their relevance scores simultaneously. 
Our initial experiments show that this batch design outperforms the one-by-one approach (i.e., rating the relevance score of each historical SV individually), 
possibly because it is more intuitive to determine the relative order of relevance than to maintain consistency in assigning absolute scores.
Refer to online appendix~\cite{replication_submit} for the detailed prompt.

\vspace{-0.1cm}
\section{Experiment Setup} \label{sec:exp_setup}

\subsection{Data Collection} \label{subsec:data_collection}
We follow existing studies~\cite{pan2024towards, pan2022automated, ponta2019dataset} to use GitHub issue reports referenced by CVE items 
as a proxy of SV reports (\sir).
The detailed data preparation steps are as follows:

\noindent \textbf{STEP1: Collect SV information.}
We first collect all CVE records from two widely recognized SV databases, i.e., NVD~\cite{NVD_website} and OSV~\cite{OSV_database}, respectively.
Notably, OSV is a well-known open-source software (OSS) SV database developed by Google.
Considering that NVD and OSV may offer different SV metadata (e.g., CVSS, CWE, and external references), 
we use OSV solely as a supplementary source to enrich external references, while extracting all other metadata from NVD to ensure consistency.
Using the CVE ID as the unified SV identifier, we merge the external references from both databases and extract GitHub issue links using regular expressions.

\noindent \textbf{STEP2: Collect \sir information.}
Based on the extracted issue links, we crawl issue data (e.g., title, body, creation date) from GitHub.
Then, we associate each crawled issue data with its corresponding CVE record and label it using the CVSS v3 metrics provided by the NVD.

\noindent \textbf{STEP3: Data cleaning.}
We exclude the following \sir data:
\ding{182}~\sirs whose corresponding CVE records lack valid CVSS v3 metrics (i.e., labels).
\ding{183}~\sirs from GitHub projects with less than 100 stars.
These projects are not widely used and are thus not representative.
Besides, we observe that issues of these projects are likely to be of low quality.
\ding{184}~\sirs created after the disclosure date of the corresponding CVE records.
These \sirs are outside the scope of early assessment, i.e., most of them are used to track the remediation process of disclosed SVs rather than initially reporting newly discovered SVs.
Besides, these \sirs may reference information from the disclosed SVs, leading to potential data leakage.
\ding{185}~\sirs correspond to multiple CVEs.
These \sirs are tangled (i.e., reporting multiple SVs simultaneously), which could confuse the assessment approach.

Finally, we collect \datasetsize \sirs from \datasetsizeproject projects.
For each project, we consider only its primary programming language as identified by the GitHub API. 
Our dataset spans 54 different programming languages, covering all of GitHub’s top-10 languages~\cite{github_top_languages}. 
Our dataset also includes 159 distinct CWE categories, encompassing the entire MITRE CWE Top-25 list~\cite{cwe_top25}. 
Detailed distributions of programming languages and CWE categories are provided in the online appendix~\cite{replication_submit}. 
Collectively, these statistics demonstrate that our dataset is highly representative.

\subsection{Experiment Setting}

\noindent \textbf{Implementation Details.}
Except for the dedicated assessment LLM (highlighted by the purple box in Figure~\ref{fig:approach_overview}), all other agents in \appname are powered by Llama-3.3-70B~\cite{llama_3.3} in our implementation.
The default temperature is set to 0.1 to ensure a consistent output.
We fine-tune Llama-3.1-8B~\cite{llama-3.1-8b} to develop the dedicated assessment LLM.
During the SFT stage, we set the learning rate to $1e^{-5}$ and train the model for two epochs.
In the subsequent RL stage, the learning rate is reduced to $1e^{-6}$ and training proceeds for two epochs.
For GRPO, we configure the rollout number to five, meaning that five outputs are sampled for each prompt.

\noindent \textbf{Baselines.}
In experiments, we include the following baselines.
    \ding{182}~\emph{ML-based baselines}. ML classifiers are widely adopted for automated SV assessment~\cite{iannone2024early, le2022survey, le2022use}. 
    Recent studies~\cite{iannone2024early, le2024mitigating} show that their performance is comparable to DL baselines in SV assessment tasks.
    We follow previous studies~\cite{pan2024towards, le2021deepcva, le2019automated, le2022use} to include the following six ML baselines: Random Forest (RF), Support Vector Machine (SVM), Logistic Regression (LR), K-Nearest Neighbors (KNN), XGBoost (XGB), and Light Gradient Boosting Machine (LGBM). 
    We adopt the same configuration and follow the same hyper-parameter tuning process as in previous studies~\cite{pan2024towards, le2021deepcva, le2022use, le2019automated}.
    For detailed settings, please refer to our online appendix~\cite{replication_submit}.
    For each ML baseline, we report the test performance under the optimal hyper-parameter setting selected by performing grid search on the validation set.
    \ding{183}~\emph{DL-based baselines}. We include DeepCVA~\cite{le2021deepcva} and \proEVA~\cite{pan2024towards} as two DL-based baselines.
    Refer to Section~\ref{sec:related_work} for a brief introduction of these two approaches.
    We configure both baselines using the hyperparameters reported in the original studies~\cite{le2021deepcva,pan2024towards}.
    Since DeepCVA was originally designed to assess SVs based on SV-inducing commits, we adapt it to our report-based setting by replacing its TextCNN-based commit embedding module with a standard TextCNN encoder for SV reports, while retaining its core multi-task learning architecture, which predicts multiple CVSS metric values from a shared input representation. Although more recent commit-based assessment approaches exist, such as CAT~\cite{li2023commit}, we do not include them as additional baselines because their primary improvements over DeepCVA lie in more advanced commit representation modules that exploit code-specific information, including program dependencies and the surrounding context of code changes. Such advantages are not applicable in our report-based setting. Once their commit encoders are replaced with a report encoder, these approaches become effectively equivalent to the adapted DeepCVA at the task-design level.
    \ding{184}~\emph{LLM-based baselines}. 
    We include three advanced LLMs, i.e., Llama-3.3-70b~\cite{llama_3.3}, DeepSeek-v3~\cite{liu2024deepseek}, and GPT-4.1~\cite{gpt_4.1},  as baselines.
    We adopt the same few-shot prompt setting introduced in our preliminary study (see Section~\ref{subsec:motivation_llm_reasoning}), which has been proven to achieve the best assessment performance among the various prompt settings.

\noindent \textbf{Evaluation Metrics.}
We follow existing studies~\cite{pan2024towards, le2024mitigating, li2023prediction} to adopt classification metrics including weighted F1 and Matthews Correlation Coefficient (MCC)~\cite{gorodkin2004comparing} to measure assessment performance.
In addition to the performance on each of the eight individual CVSS metrics, we report two types of overall assessment performance:
1)~the average performance across all eight CVSS metrics,
2)~the performance in terms of severity rating, which is calculated from the predicted values of eight CVSS metrics using the official equations~\cite{cvss_v3_specification_document}.
Note that the severity performance is not directly proportional to the average performance.
This is because 
1)~the impact of different CVSS metrics on the overall severity rating varies,
2)~opposing deviations in predictions of different CVSS metrics can offset each other, potentially resulting in no change to the overall severity rating.
Thus, the average performance of the two approaches should be comparable before it becomes meaningful to further evaluate their performance in terms of severity rating.

\section{Experiment Results} \label{sec:experiment_results}
\vspace{-0.1cm}
In the experiment, we aim to answer the following RQs:
\vspace{-0.1cm}
\begin{itemize}[leftmargin=*]
    \item \textbf{How effective is \appnamebold compared to baselines for SV assessment?}
    \item \textbf{How effective are the design choices for building a dedicated assessment LLM?}  %
    \item \textbf{How effective are the design choices for enriching SV information?}
    \item \textbf{Can the evidence provided by \appnamebold help security analysts in assessing SVs?}
\end{itemize}
\vspace{-0.1cm}

\subsection{RQ1. Effectiveness of \appnamebold for SV Assessment}
\label{subsec:RQ1}

\noindent \textbf{Method.}
To verify the effectiveness of \appname in SV assessment, we compare its performance with ML, DL, and LLM baselines using the dataset collected in Section~\ref{subsec:data_collection}.
We divide the data set into train, validation, and test sets with a ratio of 8: 1: 1.
Specifically, the dataset is split chronologically following existing works~\cite{le2021deepcva, pan2024towards, iannone2024early}, i.e., the assessment approach is trained (if necessary) using historical SVs and tested to assess future SVs.
The time-aware split has proven to be crucial for ensuring accurate and reliable evaluations of SV-related prediction tasks~\cite{falessi2020need, jimenez2019importance}.

\begin{table*}[tbp]
  \centering
  \caption{The performance comparison between \appname and baselines for SV assessment}
  \vspace{-0.3cm}
  \scalebox{0.7}{
    \begin{tabular}{c|c|cccccc|cc|ccc|c}
    \toprule
    \multirow{2}[4]{*}{\textbf{\tabincell{c}{CVSS\\Metric}}} & \multirow{2}[4]{*}{\textbf{Metrics}} & \multicolumn{12}{c}{\textbf{Method}} \\
\cmidrule{3-14}          &       & \multicolumn{1}{c}{\textbf{RF}} & \multicolumn{1}{c}{\textbf{SVM}} & \multicolumn{1}{c}{\textbf{LR}} & \multicolumn{1}{c}{\textbf{KNN}} & \multicolumn{1}{c}{\textbf{XGB}} & \multicolumn{1}{c|}{\textbf{LGBM}} & \multicolumn{1}{c}{\textbf{DeepCVA}} & \multicolumn{1}{c|}{\textbf{proEVA}} & \multicolumn{1}{c}{\textbf{Llama}} & \multicolumn{1}{c}{\textbf{DeepSeek}} & \multicolumn{1}{c|}{\textbf{GPT}} & \appnamebold \\
    \midrule
    \multirow{2}[2]{*}{\textbf{AV}} & \textbf{F1} & 0.727  & 0.788  & 0.782  & 0.767  & 0.758  & 0.744  & 0.796  & 0.790  & 0.759  & 0.790  & 0.665  & \textbf{0.856 } \\
          & \textbf{MCC} & 0.269  & 0.391  & 0.375  & 0.327  & 0.308  & 0.257  & 0.451  & 0.406  & 0.473  & 0.531  & 0.422  & \textbf{0.615 } \\
    \midrule
    \multirow{2}[1]{*}{\textbf{AC}} & \textbf{F1} & \textbf{0.972 } & \textbf{0.974 } & 0.970  & \textbf{0.972 } & \textbf{0.974 } & \textbf{0.977 } & \textbf{0.972 } & \textbf{0.982 } & \textbf{0.977 } & 0.951  & 0.967  & \textbf{0.979 } \\
          & \textbf{MCC} & 0.000  & 0.261  & 0.096  & 0.000  & 0.159  & 0.281  & 0.119  & \textbf{0.497 } & 0.389  & 0.236  & 0.322  & 0.405  \\
    \midrule
    \multirow{2}[1]{*}{\textbf{PR}} & \textbf{F1} & 0.747  & 0.815  & 0.801  & 0.769  & 0.812  & 0.809  & 0.809  & 0.792  & \textbf{0.831 } & 0.821  & 0.795  & \textbf{0.830 } \\
          & \textbf{MCC} & 0.207  & 0.401  & 0.352  & 0.300  & 0.395  & 0.389  & 0.399  & 0.321  & \textbf{0.466 } & 0.423  & 0.386  & \textbf{0.464 } \\
    \midrule
    \multirow{2}[1]{*}{\textbf{UI}} & \textbf{F1} & 0.742  & 0.768  & 0.789  & 0.613  & 0.764  & 0.770  & 0.788  & 0.813  & 0.659  & 0.705  & 0.769  & \textbf{0.877 } \\
          & \textbf{MCC} & 0.486  & 0.517  & 0.556  & 0.328  & 0.503  & 0.515  & 0.553  & 0.607  & 0.298  & 0.364  & 0.516  & \textbf{0.736 } \\
    \midrule
    \multirow{2}[1]{*}{\textbf{S}} & \textbf{F1} & 0.939  & 0.948  & 0.953  & 0.924  & 0.962  & 0.958  & 0.950  & 0.949  & 0.779  & 0.826  & 0.918  & \textbf{0.976 } \\
          & \textbf{MCC} & 0.714  & 0.750  & 0.778  & 0.637  & 0.818  & 0.801  & 0.767  & 0.763  & -0.056  & 0.325  & 0.651  & \textbf{0.888 } \\
    \midrule
    \multirow{2}[1]{*}{\textbf{C}} & \textbf{F1} & 0.740  & 0.684  & 0.740  & 0.594  & 0.711  & 0.769  & 0.758  & 0.737  & 0.708  & 0.718  & 0.664  & \textbf{0.809 } \\
          & \textbf{MCC} & 0.572  & 0.482  & 0.567  & 0.332  & 0.529  & 0.618  & 0.599  & 0.567  & 0.551  & 0.549  & 0.529  & \textbf{0.686 } \\
    \midrule
    \multirow{2}[1]{*}{\textbf{I}} & \textbf{F1} & 0.709  & 0.712  & 0.708  & 0.555  & 0.692  & 0.689  & 0.734  & 0.754  & 0.700  & 0.683  & 0.725  & \textbf{0.802 } \\
          & \textbf{MCC} & 0.512  & 0.520  & 0.514  & 0.263  & 0.494  & 0.490  & 0.566  & 0.590  & 0.517  & 0.481  & 0.556  & \textbf{0.673 } \\
    \midrule
    \multirow{2}[1]{*}{\textbf{A}} & \textbf{F1} & 0.784  & 0.811  & 0.808  & 0.709  & 0.827  & 0.833  & 0.845  & 0.828  & 0.809  & 0.774  & 0.749  & \textbf{0.866 } \\
          & \textbf{MCC} & 0.507  & 0.553  & 0.548  & 0.298  & 0.596  & 0.611  & 0.643  & 0.600  & 0.577  & 0.524  & 0.499  & \textbf{0.696 } \\
    \midrule
    \rowcolor[rgb]{ .851,  .851,  .851} 
    & \textbf{F1} & 0.795  & 0.812  & 0.819  & 0.738  & 0.812  & 0.819  & 0.832  & 0.831  & 0.778  & 0.783  & 0.781  & \textbf{0.874 } \\
    \rowcolor[rgb]{ .851,  .851,  .851}       
    \multirow{-2}{*}{\textbf{Average}} & \textbf{MCC} & 0.408  & 0.484  & 0.473  & 0.311  & 0.475  & 0.495  & 0.512  & 0.544  & 0.402  & 0.429  & 0.485  & \textbf{0.646 } \\
    \midrule
    \rowcolor[rgb]{ .851,  .851,  .851} 
    & \textbf{F1} & 0.468  & 0.540  & 0.546  & 0.450  & 0.556  & 0.572  & 0.549  & 0.587  & 0.572  & 0.492  & 0.405  & \textbf{0.672 } \\
    \rowcolor[rgb]{ .851,  .851,  .851}
    \multirow{-2}{*}{\textbf{Severity}}
    & \textbf{MCC} & 0.249  & 0.304  & 0.309  & 0.154  & 0.334  & 0.359  & 0.298  & 0.363  & 0.328  & 0.218  & 0.108  & \textbf{0.490 } \\
    \bottomrule
    \end{tabular}
    \vspace{-0.5cm}
}
  \label{table:assessment_baseline}%
\end{table*}%

\noindent \textbf{Results.}
Table~\ref{table:assessment_baseline} presents the performance comparisons between \appname and baselines.
LGBM achieves the best performance among the ML baselines.
The two DL baselines (i.e., DeepCVA and \proEVA) slightly improve the performance of LGBM.
For three LLM baselines, 
their performance is worse than the ML- and DL-based baselines by a large margin.
This finding is in line with our preliminary study, which verifies the necessities of domain-specific knowledge in SV assessment.

Regarding the two overall measures, our approach yields the best performance across all evaluation metrics.
For the average measure, \appname surpasses the best-performing baseline (i.e., \proEVA) by 5.3\% and 18.7\% in terms of weighted F1 and MCC, respectively.
For the severity measure, the improvements are 14.4\% and 35.2\%, respectively.
Notably, the improvements in MCC are considerably more pronounced than those in weighted F1, which can be attributed to the imbalanced class distributions across CVSS metrics~\cite{le2021deepcva}. 
While weighted F1 is primarily driven by performance on majority classes, MCC measures the global consistency between predictions and ground truth across all classes, and thus more effectively captures improvements in minority-class performance and inter-class discrimination. 
The substantial gains in MCC therefore indicate that \appname excels at distinguishing between classes.

Regarding individual CVSS metrics, \appname outperforms \proEVA on all metrics except \emph{Attack Complexity}. 
The most substantial improvements are observed on \emph{Confidentiality}, \emph{Attack Vector}, \emph{User Interaction}, and \emph{Integrity}, with F1 gains of 9.7\%, 8.3\%, 7.9\%, and 6.5\%, respectively.
In contrast, improvements on the remaining metrics, namely \emph{Privileges Required}, \emph{Availability}, and \emph{Scope}, are less pronounced, at 4.7\%, 4.7\%, and 2.9\% in F1, respectively.
This is primarily because \proEVA already achieves strong performance on these metrics. 
For example, \proEVA attains a high F1 of 0.949 on the \emph{Scope} metric, while \appname further improves it to 0.976.
Overall, these results confirm the consistent advantages of \appname over existing approaches across CVSS metrics.

\vspace{-0.1cm}
\find{{\rm \bf RQ1:} 
\appname significantly outperforms baselines in SV assessment, consistently improving performance across CVSS metrics, with especially large gains on those metrics the baselines are weaker.
}
\vspace{-0.1cm}

\subsection{RQ2. Effectiveness of Designs for Training a Dedicated Assessment LLM}
\label{subsec:RQ2}
\noindent \textbf{Method.}
We further verify the effectiveness of key designs for finetuning LLM to effectively inject assessment-specific knowledge.
We compare the performance of \appname with several variants, each lacking one of the key designs:
\begin{itemize}[leftmargin=*]

\item The \emph{w/o finetune} variant is designed to examine the necessity of finetuning when building a specialized assessment LLM.
We remove fine-tuning, which embeds historical SV data into the model’s internal knowledge (i.e., its parameters). Instead, we use historical SV data (i.e., the training and validation sets) to construct an external knowledge base, which the vanilla LLM accesses through a retrieval-augmented generation (RAG) process.
Note that the historical SV data is enriched with evidence prepared using the methods described in Section~\ref{subsec:approach_finetune}.
Our initial tries indicate that including evidence, in addition to the label metric value, in the demonstration examples significantly improves performance.
We adopt the same \emph{few-shot} prompt setting as in our preliminary study, 
which has been proven to be effective.
Specifically, we retrieve the top 10 historical SVs using a hybrid retriever (see Section~\ref{subsec:approach_retrieve}) and select one example per candidate metric value to construct the few-shot demonstrations.
We use Llama-3.3-70b for this variant instead of Llama-3.1-8b used to build the dedicated assessment LLM, which offers stronger off-the-shelf ability.

\item The \emph{w/o reason} variant is designed to examine whether explicitly generating a reasoning process improves the assessment performance. For this variant, we finetune the model to directly predict the metric value without first producing an explicit reasoning trajectory.

\item The \emph{w/o RL} variant is designed to assess the effectiveness of applying reinforcement learning (RL) on top of conventional SFT in enhancing the model’s reasoning capability. Specifically, we report the performance of using SFT alone, without the subsequent GRPO training (see Section~\ref{subsec:approach_finetune}).

\end{itemize}

\begin{table}[tbp]
  \centering
  \caption{The results for assessing the effectiveness of designs for training a dedicated assessment LLM}
   \vspace{-0.2cm}
  \scalebox{0.8}{
    \begin{tabular}{clcccccccccc}
    \toprule
    \textbf{Metrics} & \multicolumn{1}{c}{\textbf{Approach}} & \textbf{AV} & \textbf{AC} & \textbf{PR} & \textbf{UI} & \textbf{S} & \textbf{C} & \textbf{I} & \textbf{A} & \cellcolor[rgb]{ .851,  .851,  .851}\textbf{Average} & \cellcolor[rgb]{ .851,  .851,  .851}\textbf{Severity} \\
    \midrule
    \multirow{4}[4]{*}{\textbf{F1}} & \textbf{w/o finetune} & 0.795  & 0.933  & \textbf{0.830 } & 0.687  & 0.805  & 0.760  & 0.713  & 0.832  & \cellcolor[rgb]{ .851,  .851,  .851}0.794  & \cellcolor[rgb]{ .851,  .851,  .851}0.584  \\
          & \textbf{w/o reason} & 0.728  & \textbf{0.972 } & 0.807  & 0.830  & \textbf{0.975 } & 0.786  & 0.768  & 0.849  & \cellcolor[rgb]{ .851,  .851,  .851}0.839  & \cellcolor[rgb]{ .851,  .851,  .851}0.566  \\
          & \textbf{w/o RL} & 0.797  & \textbf{0.975 } & \textbf{0.822 } & 0.821  & \textbf{0.971 } & 0.784  & 0.748  & \textbf{0.866 } & \cellcolor[rgb]{ .851,  .851,  .851}0.848  & \cellcolor[rgb]{ .851,  .851,  .851}0.617  \\
\cmidrule{2-12}          & \appnamebold & \textbf{0.856 } & \textbf{0.979 } & \textbf{0.830 } & \textbf{0.877 } & \textbf{0.976 } & \textbf{0.809 } & \textbf{0.802 } & \textbf{0.866 } & \cellcolor[rgb]{ .851,  .851,  .851}\textbf{0.874 } & \cellcolor[rgb]{ .851,  .851,  .851}\textbf{0.672 } \\
    \midrule
    \multirow{4}[4]{*}{\textbf{MCC}} & \textbf{w/o finetune} & 0.541  & 0.250  & 0.448  & 0.403  & 0.074  & 0.620  & 0.538  & 0.578  & \cellcolor[rgb]{ .851,  .851,  .851}0.431  & \cellcolor[rgb]{ .851,  .851,  .851}0.351  \\
          & \textbf{w/o reason} & 0.454  & 0.000  & 0.388  & 0.636  & \textbf{0.881 } & 0.657  & 0.623  & 0.659  & \cellcolor[rgb]{ .851,  .851,  .851}0.537  & \cellcolor[rgb]{ .851,  .851,  .851}0.337  \\
          & \textbf{w/o RL} & 0.490  & 0.226  & 0.423  & 0.614  & 0.865  & 0.646  & 0.584  & \textbf{0.704 } & \cellcolor[rgb]{ .851,  .851,  .851}0.569  & \cellcolor[rgb]{ .851,  .851,  .851}0.397  \\
\cmidrule{2-12}          & \appnamebold & \textbf{0.615 } & \textbf{0.405 } & \textbf{0.464 } & \textbf{0.736 } & \textbf{0.888 } & \textbf{0.686 } & \textbf{0.673 } & \textbf{0.696 } & \cellcolor[rgb]{ .851,  .851,  .851}\textbf{0.646 } & \cellcolor[rgb]{ .851,  .851,  .851}\textbf{0.490 } \\
    \bottomrule
    \end{tabular}
}
  \label{table:assessment_ablation}
  \vspace{-0.2cm}
\end{table}

\noindent \textbf{Results.}
Table~\ref{table:assessment_ablation} presents the performance comparisons between \appname and variants.
Regarding the average measure, \appname outperforms the \emph{w/o finetuning} variant by 10.1\%, and 49.6\% in terms of weighted F1 and MCC, respectively.
Regarding the severity measure, the improvements are 15.1\% and 39.6\%, respectively.
Besides, \appname consistently outperforms the \emph{w/o finetuning} variant on every specific CVSS metric.
The most significant enhancements are observed on \emph{User Interaction} and \emph{Scope}, with improvements of 27.6\% and 21.3\% in terms of weighted F1, respectively.
For the assessment of these CVSS metrics, the domain knowledge is the most demanding. 
These results verify the necessity of injecting assessment-specific knowledge through finetuning (i.e., a dense solution), which is more efficient than RAG (i.e., a sparse solution).
Compared with the \emph{w/o reason} variant, \appname consistently achieves better performance on both overall measures and all individual CVSS metrics.
Specifically, on the average measure, \appname improves weighted F1 and MCC by 4.2\% and 20.1\%, respectively. 
These results suggest that explicitly learning the reasoning process enables LLMs to more accurately determine the correct metric values.
Furthermore, \appname also significantly outperforms the \emph{w/o RL} variant across both overall measures and all individual CVSS metrics. 
For the average measure, the gains reach 3.1\% in weighted F1 and 13.5\% in MCC, demonstrating that applying RL on top of conventional SFT further enhances the model’s intrinsic reasoning capability.
In addition, \appname exhibits similar performance gains over both the \emph{w/o reason} and \emph{w/o RL} variants across individual CVSS metrics.
The largest improvements are observed for \emph{Attack Vector}, \emph{User Interaction}, and \emph{Integrity}, suggesting that effective reasoning over surface-level SV information is particularly critical for accurately assessing these metrics.

\vspace{-0.1cm}
\find{{\rm \bf RQ2:} 
Injecting assessment-specific knowledge through fine-tuning is more effective than a RAG-based solution. 
Explicitly learning the reasoning process for determining metric values further improves assessment performance. 
Applying RL on top of conventional SFT enhances the model’s intrinsic reasoning capability.
}
\vspace{-0.1cm}

\subsection{RQ3. Effectiveness of Designs for Enriching SV Information}
\label{subsec:RQ3}
\noindent \textbf{Method.}
We further verify the effectiveness of the designs for enriching SV information by processing rich text content in \sirs and incorporating information of the affected repository.
Specifically, we compare the performance of \appname with three variants, i.e., \emph{w/o code}, \emph{w/o screen}, and \emph{w/o repo}, each of which removes one type of enriched SV information.

\begin{table}[tbp]
  \centering
  \caption{The results for assessing the effectiveness of designs for enriching SV information}
  \vspace{-0.2cm}
  \scalebox{0.8}{
     \begin{tabular}{clcccccccccc}
    \toprule
    \textbf{Metrics} & \multicolumn{1}{c}{\textbf{Approach}} & \textbf{AV} & \textbf{AC} & \textbf{PR} & \textbf{UI} & \textbf{S} & \textbf{C} & \textbf{I} & \textbf{A} & \cellcolor[rgb]{ .851,  .851,  .851}\textbf{Average} & \cellcolor[rgb]{ .851,  .851,  .851}\textbf{Severity} \\
    \midrule
    \multirow{4}[4]{*}{\textbf{F1}} & \textbf{w/o code} & 0.826  & \textbf{0.979 } & \textbf{0.838 } & 0.854  & \textbf{0.973 } & 0.800  & 0.791  & 0.797  & \cellcolor[rgb]{ .851,  .851,  .851}0.857  & \cellcolor[rgb]{ .851,  .851,  .851}0.634  \\
          & \textbf{w/o screen} & 0.833  & \textbf{0.978 } & 0.813  & 0.865  & \textbf{0.973 } & \textbf{0.821 } & \textbf{0.805 } & 0.854  & \cellcolor[rgb]{ .851,  .851,  .851}\textbf{0.868 } & \cellcolor[rgb]{ .851,  .851,  .851}0.658  \\
          & \textbf{w/o repo} & 0.829  & \textbf{0.979 } & 0.822  & 0.849  & \textbf{0.975 } & 0.802  & \textbf{0.799 } & \textbf{0.863 } & \cellcolor[rgb]{ .851,  .851,  .851}\textbf{0.865 } & \cellcolor[rgb]{ .851,  .851,  .851}0.646  \\
\cmidrule{2-12}          & \appnamebold & \textbf{0.856 } & \textbf{0.979 } & \textbf{0.830 } & \textbf{0.877 } & \textbf{0.976 } & 0.809  & \textbf{0.802 } & \textbf{0.866 } & \cellcolor[rgb]{ .851,  .851,  .851}\textbf{0.874 } & \cellcolor[rgb]{ .851,  .851,  .851}\textbf{0.672 } \\
    \midrule
    \multirow{4}[4]{*}{\textbf{MCC}} & \textbf{w/o code} & 0.509  & \textbf{0.405 } & \textbf{0.484 } & 0.684  & 0.873  & 0.672  & 0.652  & 0.564  & \cellcolor[rgb]{ .851,  .851,  .851}0.605  & \cellcolor[rgb]{ .851,  .851,  .851}0.433  \\
          & \textbf{w/o screen} & 0.546  & 0.328  & 0.399  & 0.710  & 0.873  & \textbf{0.704 } & \textbf{0.677 } & 0.665  & \cellcolor[rgb]{ .851,  .851,  .851}0.613  & \cellcolor[rgb]{ .851,  .851,  .851}0.470  \\
          & \textbf{w/o repo} & 0.517  & \textbf{0.405 } & 0.431  & 0.678  & \textbf{0.881 } & 0.676  & \textbf{0.668 } & \textbf{0.693 } & \cellcolor[rgb]{ .851,  .851,  .851}0.619  & \cellcolor[rgb]{ .851,  .851,  .851}0.451  \\
\cmidrule{2-12}          & \appnamebold & \textbf{0.615 } & \textbf{0.405 } & 0.464  & \textbf{0.736 } & \textbf{0.888 } & 0.686  & \textbf{0.673 } & \textbf{0.696 } & \cellcolor[rgb]{ .851,  .851,  .851}\textbf{0.646 } & \cellcolor[rgb]{ .851,  .851,  .851}\textbf{0.490 } \\
    \bottomrule
    \end{tabular}
}
  \label{table:assessment_ablation_richtext}
  \vspace{-0.2cm}
\end{table}%

\noindent \textbf{Results.}
Table~\ref{table:assessment_ablation_richtext} presents the performance comparisons between \appname and the variants.
\appname achieves the best overall performance.
Preprocessing embedded code snippets yields the most significant benefits among the three types of enriched SV information. For the severity measure, this results in improvements of 8.5\% and 13.4\% in terms of weighted F1 and MCC, respectively. This is likely because long and noisy code snippets can distract LLMs from the key information needed for accurate assessment.
Incorporating screenshot information improves the severity measure by 2.2\% and 4.3\% in terms of weighted F1 and MCC, respectively. 
For individual CVSS metrics, incorporating screenshot information consistently improves performance across all metrics except \emph{Confidentiality}, with the largest gains observed for \emph{Attack Vector} and \emph{Privileges Required}, achieving MCC improvements of 12.7\% and 16.3\%, respectively.
Incorporating information about the vulnerable project yields consistent improvements on both overall measures and all individual CVSS metrics, achieving gains of 4.4\% in weighted F1 and 8.9\% in MCC for the severity measure. 
Collectively, these results confirm the effectiveness of processing rich text content in \sirs and integrating vulnerable project information for accurate SV assessment.

\vspace{-0.1cm}
\find{{\rm \bf RQ3:}
Incorporating rich text content from \sirs and information about the vulnerable project enhances assessment performance.
}
\vspace{-0.1cm}

\subsection{RQ4. Necessities and Effectiveness of the Evidence Provided by \appnamebold in Practice}
\label{subsec:RQ4}
\vspace{-0.1cm}
\noindent \textbf{Method.}
In the previous RQs, we demonstrate that \appname achieves accurate SV assessment. We further examine the necessity and effectiveness of providing supporting evidence (i.e., reasoning processes) for the assigned metric values. 
As discussed in Section~\ref{subsec:motivation_challenge}, prior studies focus primarily on assessment accuracy while overlooking the need to explain how assessment results are derived.
We therefore conduct a user study to evaluate how the supporting evidence provided by \appname assists analysts in determining assessment results.

\noindent \emph{\underline{Tasks:}} 
We randomly select 40 SVs from our test set, resulting in a total of 320 metric-wise assessment tasks.
For each SV, we provide participants with
1) the SV report and relevant CVSS standard,
2) the value assigned and the analysis generated by \appname,
3) the information of similar historical SVs, including CVE-ID, report, and the analysis generated by \appname. 
We then ask participants to answer the following questions for each of the eight CVSS metrics.
\vspace{-0.1cm}
\begin{enumerate}[label=\arabic*), leftmargin=*]
    \item Choose the appropriate metric value.
    \item Rate (using 5-point Likert scale) the usefulness of the provided evidence.
    \item Rate the factuality (i.e., the evidence is grounded in the given SV information), relevance (i.e., the evidence directly supports the assigned metric value), and completeness (i.e., the evidence covers key metric-relevant details from the given SV information) of the provided evidence.
    
    \item Rate the usefulness of the provided historical SV information in determining the metric value. Participants can skip Q4 if they think the information of the SV itself is enough to make decision.
\end{enumerate}
\vspace{-0.1cm}

\noindent \emph{\underline{Participants:}}
We invite 8 security practitioners as participants.
All of them have three to five years of experience in the software security domain and are familiar with the CVSS standard.

\noindent \textbf{Results.}
Among the 284 cases where \appname predicts the correct metric value, participants acknowledge the usefulness of the provided evidence (rating 4/5) in 275 (96.8\%) cases.
For these cases, participants generally rate the provided evidence with high factuality, relevance, and completeness (average rating all above 4.6).
We further investigate 54 cases where participants think the provided evidence is less useful (rating$\le$4).
We observe that the ratings for factuality and completeness remain high, but downgrade on relevance.
Additionally, we also observe a much higher ratio of historical SV information being examined.
We interview participants and confirm that, in these cases, 
they feel less confident making decisions based solely on the information of the SV itself,
prompting them to further examine the provided historical SV information.
Regarding the usefulness of the historical SV information provided by \appname, 
among the 89 cases examined by participants, 78 (87.6\%) is rated with 4/5.
For the left cases where the retrieved historical SVs are less useful, 
we find that this is mostly because the query SV is closely tied to the project-specific attributes.

Among the 36 cases where \appname makes incorrect predictions, participants still select the correct metric value in 12 cases, suggesting that they are unlikely to be misled. 
In these cases, the evidence retains similarly high factuality but receives lower relevance and completeness ratings than evidence accompanying correct predictions. 
This indicates that the evidence given by \appname remains fact-based, but can be irrelevant or miss the key information related to the correct metric value.
Interviews and manual inspection show that \appname generally analyzes the SV accurately but either misinterprets the CVSS criteria or provides overly general evidence that does not support a specific value.
We further investigate the left 24 cases where participants fail to select the correct metric value.
In 18 cases, the publicly available SV information is insufficient to determine a unique CVSS metric value, or the value assigned by NVD may be inappropriate~\cite{synk_different_cvss_1, synk_different_cvss_2}.
For example, CVE-2023-39138~\cite{zipfoundation_issue_282} is a path traversal SV caused by insufficient validation of whether a symlink resides within the directory during ZIP file extraction.
The impact of this SV mostly concerns \emph{Confidentiality} and \emph{Integrity}, as attackers can assess and modify sensitive information via crafted symlinks.
The SV report does not mention any impact on \emph{Availability}.
However, NVD assigns \emph{High} for \emph{Availability} metric, while Synk~\cite{CVE-2023-39138_synk} (a commercial SV database) and \appname assign \emph{None}.

\vspace{-0.2cm}
\find{{\rm \bf RQ4:}
Evidence provided by \appname can effectively help analysts in determine the assessment results.
Participants are unlikely to be misled by \appname.
}
\vspace{-0.2cm}

\section{Discussion} \label{subsec:discussion}
\subsection{Generalizability of \appnamebold Across Different Underlying LLMs}

From a methodological perspective, \appname's two-stage fine-tuning framework, which combines SFT and RL (see Section~\ref{subsec:approach_finetune}), is not tied to any particular foundation model. To further evaluate its generalizability, we construct the assessment LLM using Qwen3-8B and compare it with the original implementation based on Llama-3.1-8B. As shown in Table~\ref{table:foundation_model_comparison}, \appname achieves broadly comparable performance with the two underlying LLMs across all evaluation metrics. These results suggest that \appname generalizes effectively across foundation models of similar scale.

\begin{table}[t]
\centering
\caption{Performance comparison across different underlying LLMs}
\vspace{-0.1cm}
\label{table:foundation_model_comparison}
\begin{tabular}{lcccc}
\toprule
\textbf{Base LLM}
& \textbf{F1 (Average)}
& \textbf{F1 (Severity)}
& \textbf{MCC (Average)}
& \textbf{MCC (Severity)} \\
\midrule
Llama-3.1-8B & 0.874 & 0.672 & 0.646 & 0.490 \\
Qwen3-8B     & 0.872 & 0.665 & 0.622 & 0.490 \\
\bottomrule
\end{tabular}
\end{table}

\subsection{Computational Overhead of \appnamebold}

\begin{table*}[t]
\centering
\caption{Computational overhead for processing a single SV report.}
\vspace{-0.1cm}
\small
\begin{tabular}{lccc}
\toprule
\textbf{Stage}
& \textbf{\tabincell{c}{Information\\Enrichment}} 
& \textbf{\tabincell{c}{Assessment\\(Fine-tuned Llama-3.1-8B)}}
& \textbf{\tabincell{c}{Historical SV\\Retrieval}} \\
\midrule
\textbf{Time (s)}
& 14.77 (screenshot) + 5.46 (code + repo)
& 5.65
& 84.8 \\
\textbf{Cost (\$)}
& 0.0013 (screenshot) + 0.0008 (code + repo)
& /
& 0.0306 \\
\textbf{\# Tokens}
& 2,166 (screenshot) + 952 (code + repo)
& 12,718
& 34,733 \\
\bottomrule
\end{tabular} \label{table:computation}
\end{table*}

Table~\ref{table:computation} presents the average inference latency and cost for processing a single SV report (including the analysis of all eight CVSS metrics).
The overall monetary cost is low (0.0327\$), and the latency can be easily reduced through parallelism. For instance, by increasing the batch size to eight, the latency of the assessment stage can be reduced to 1.52s. We believe \appname is highly cost-effective, largely because the LLMs used to build it (e.g., Llama-3.1-8B and Llama-3.3-70B) are open-source and of small/moderate size. 

Notably, both the latency and monetary cost incurred before the \emph{Historical SVs Retrieval} stage are extremely low (25.88s \& 0.0021\$). The relatively high costs of the \emph{Historical SVs Retrieval} stage are primarily due to the use of an Llama-3.3-70b (for each of the eight CVSS metrics) to rerank the 10 retrieved historical SVs. However, the cost of this stage can be avoided when the SV report already provides sufficient information, as demonstrated in our user study (see Section~\ref{subsec:RQ4}). We plan to refine the pipeline so that this stage is only triggered when explicitly requested by an analyst.

\subsection{Validity of Analyzing Screenshots and Code Snippets}
We randomly sample 50 screenshots and 50 code snippets and engage two annotators to independently assess the faithfulness of LLM’s analysis using 5-point likert scale. 
For screenshots, the two annotators provide identical ratings in 48 cases and the remaining two cases are discussed to reach a consensus. 
48 screenshots receive the highest score of 5.
For the remaining two cases:
\ding{182}~The LLM includes redundant information from previously analyzed screenshots rather than focusing on the current one (rated 4).
\ding{183}~LLM mistakenly identifies the calculator application launched via a remote-code-execution vulnerability as a hex editor (rated 3).
For code snippets, both annotators consistently assign a score of 5 across all 50 cases.
These results indicates that our proposed LLM agent (see Section~\ref{subsec:approach_richtext}) can faithfully analyze the rich-text content embedded in \sirs.

\subsection{Threats to Validity} \label{subsec:threats}
\noindent \textbf{Internal Validity}
We use the CVSS values provided by NVD as labels.
However, the values assigned by NVD may be inaccurate~\cite{synk_different_cvss_1, synk_different_cvss_2}.
This is primarily because the SV assessment process is highly dependent on the information available to the analyst, the analyst’s familiarity with the SV context (e.g., the affected projects), and their expertise and experience~\cite{synk_cvss_intro, allodi2018identifying, spring2018towards}.
It is further demonstrated by the fact that different SV databases may assign varying CVSS values to the same SV.
For example, NVD and Synk assign different CVSS scores to CVE-2023-29578 (i.e., 8.8 and 5.5).
However, 
NVD CVSS scores are widely recognized by both industry and government as the standard for software vulnerability assessment~\cite{cisa_bod, fedramp_report, pci_cvss}.
Another potential threat is pretraining data contamination, as the knowledge cutoff of the LLM used to build EAVA (i.e., Llama-3.1-8B) postdates the disclosure dates of some SVs in our test set. To prevent leakage between our task-specific training and test data, we follow prior work and adopt a chronological split (see Section~\ref{subsec:RQ1}), such that EAVA is trained on historical SVs and evaluated on future SVs. We further note that the knowledge cutoffs of the three LLM baselines examined in our preliminary study and experiments (Llama-3.3-70B, DeepSeek-V3, and GPT-4.1) also overlap with part of our test data. Nevertheless, these models perform substantially worse than EAVA, providing evidence that potential pretraining contamination is unlikely to affect the validity of our findings.

\noindent \textbf{External Validity}
We build our \sir dataset using only the SV data from NVD and report data from GitHub, which may not represent all SV reports.
Future works should investigate the generalizability of \appname regarding different \sir proxies, i.e., a different SV database (e.g., Synk) or a different source of SV reports (e.g., Bugzilla).
Another threat is that we only consider CVSS as the standard for SV assessment. 
There are other assessment standards, e.g., Exploit Prediction Scoring System (EPSS) is a dedicated standard for estimating the likelihood of SVs being exploited in the wild~\cite{epss}.

\vspace{-1mm}
\section{Conclusion} \label{sec:conclusion}
In this paper, we propose \appname, a novel framework that effectively leverages LLMs to perform SV assessment while providing supporting evidence. 
Through extensive evaluations with mainstream LLMs, we demonstrate that off-the-shelf LLMs struggle to produce accurate assessments in the absence of domain-specific knowledge. 
At the core of \appname is a dedicated assessment LLM, built through large-scale trajectory annotation and a two-stage fine-tuning paradigm that combines SFT and RL to inject assessment-specific knowledge.
\appname first employs specialized LLM agents to process rich text content in \sirs and incorporate information about vulnerable projects. 
The dedicated assessment LLM then reasons over the enriched SV information to determine assessment results. 
Finally, similar historical SVs are retrieved as supplementary evidence to support assessment decisions.
Experimental results show that \appname significantly outperforms baselines and validate the effectiveness of our key design choices. 
Moreover, a user study confirms both the necessity and usefulness of the evidence provided by \appname in facilitating SV assessment.

\section{Data Availability}
The replication package of our work is publicly available at~\cite{replication_submit}.

\section{Acknowledgments}
This research/project is supported by the National Key R\&D Program of China (No. 2024YFB4506400) and Sponsored by CCF-Huawei Populus Grove Fund.

\balance
\bibliographystyle{ACM-Reference-Format}
\bibliography{reference}

\end{document}